\documentclass[useAMS,usenatbib]{mn2e}
\usepackage{graphicx}

\title[NGC 5824 intrinsic abundance spread]
{NGC 5824: a luminous outer halo globular cluster with an intrinsic abundance spread}
\author[Da Costa, Held \& Saviane]
{G. S. Da Costa,$^1$ E. V. Held,$^2$ and I. Saviane$^3$\\
$^1$Research School of Astronomy and Astrophysics, Australian National 
University, Canberra, ACT 0200, Australia\\
$^2$INAF, Osservatorio Astronomico di Padova, vicolo Osservatorio 5, 35122 Padova, Italy\\
$^3$European Southern Observatory, Alonso de Cordova 3107, Santiago, Chile}
    
\begin{document}

\maketitle

\begin{abstract}

We present a detailed study of the strengths of the calcium triplet absorption lines in the spectra of a large
sample of red giant members of the luminous outer Galactic halo globular cluster NGC~5824.  The 
spectra were obtained with the FORS2 and GMOS-S multi-object spectrographs at the VLT and the Gemini-S 
telescope, respectively.  By comparing the line strengths of the NGC~5824 stars with those for red giants 
in clusters with well established abundances, we conclude that there is an intrinsic abundance  dispersion in
NGC~5824 characterized by an inter-quartile range in [Fe/H] of 0.10 dex and a total range of $\sim$0.3 dex.
As for $\omega$~Cen and M22, the abundance distribution shows a steep rise on the metal-poor side and a
shallower decline on the metal-rich side.  There is also some indication that the distribution is not
unimodal with perhaps 3 distinct abundance groupings present. 
NGC~5824 has a further unusual characteristic: the outer surface density profile shows no signs of a 
tidal cutoff.  Instead the profile has a power-law distribution with cluster stars detected to a radius 
exceeding 400 pc.
We postulate that NGC~5824 may be the remnant nuclear star cluster of a now disrupted dwarf galaxy accreted
during the formation of the Galaxy's halo.  We further speculate that the presence of an intrinsic [Fe/H] spread is
the characteristic that distinguishes former nuclear star clusters from other globular clusters.

\end{abstract}


\section{Introduction} \label{Intro} 

On the basis of many detailed spectroscopic and photometric studies carried out over more than a 
decade, we can now assert with some conviction that most, if not all, of the globular clusters associated 
with the Milky Way galaxy are not the simple stellar populations they were once considered to be.  
The evidence to support this assertion can be found in, for example, the sodium-oxygen abundance 
anti-correlation that pervades the stars in essentially all Galactic globular clusters
\citep[see the recent review of][]{GCB12}.  While the exact nature of the ``polluters'' that give rise to the
chemical anomalies remains uncertain, it is clear that the production of the anomalies is 
intimately connected to the formation process of the cluster, and not to any process involving the current 
generation of cluster stars.

Nevertheless, despite the ubiquity of the abundance inhomogeneities involving the light elements 
C, N, O, Na, Mg and Al, it remains the case that most globular clusters are chemically homogeneous
with respect to the heavier elements such as Fe and Ca \citep[e.g.,][]{CB10c}, elements whose 
nucleosynthesis lies with supernovae.
The classic exception to this general situation is the stellar system $\omega$~Cen, which
has been known to have an unusual stellar population for almost four decades.  Specifically, the stars in
$\omega$~Cen possess a large range in iron abundance together with distinct element-to-iron
abundance ratios \citep[e.g.,][and the references therein]{JP10,AM12}.  These unusual characteristics 
have led to the suggestion that the stellar system $\omega$~Cen is ``special'', in that it is the nuclear
remnant of a now disrupted dwarf galaxy \citep[e.g.,][]{KCF93}.  
Chemical evolution models of such systems have been
moderately successful in reproducing the observed properties of the cluster \citep[e.g.,][]{DM07,DM10,
MS07} and dynamical models of the tidal disruption process
 \citep[e.g.,][]{BF03} have also shown that plausible scenarios
exist in which the nuclear remnant can end up in an orbit similar to the tightly bound and retrograde
orbit of the current cluster.  
The strongest evidence in support of this hypothesis for the origin of $\omega$~Cen, however, comes 
from the recent discovery of field stars that show unusual chemical abundance ratios similar to 
cluster member stars \citep{WdB10, SM12}.

It is now recognised, however, that $\omega$~Cen is not the only Galactic globular cluster with an
internal [Fe/H] abundance spread, although it remains the object with the largest star-to-star [Fe/H] range.
For example, the nuclear star cluster of the Sagittarius dwarf galaxy, M54, was first suggested
to have an intrinsic abundance spread, $\sigma_{int}$([Fe/H]), of $\sim$0.16 dex by \citet{SL95}
on the basis of the intrinsic colour width of the cluster red giant branch.  
Subsequent analyses based on intermediate- and high-resolution spectra
of a significant number of M54 red giants have confirmed the existence of this intrinsic spread,
yielding $\sigma_{int}$([Fe/H]) $\approx$ 0.19 dex \citep[e.g.,][and the references therein]{CB10a}.
The Sgr dwarf is currently under going tidal disruption, which when complete will leave M54 as a 
member of the Galactic halo globular cluster population, and its status as a nuclear star cluster 
of a dwarf galaxy will then no longer be obvious.  

A third cluster with an intrinsic range in [Fe/H] abundance is M22.  \citet{DC09} used 
intermediate-resolution spectra at the Ca II triplet of 41 member stars to demonstrate that there is a
broad [Fe/H] abundance distribution in this cluster qualitatively similar (but on a smaller scale) to that for
$\omega$~Cen.  The abundance distribution reveals the presence of at least two components and 
is characterised by an interquartile range in [Fe/H] of 0.24 dex \citep{DC09}.  
Similar results have been found from the high dispersion spectroscopic analyses of \citet{AM09, AM11}.  
These data show the presence of two populations in the cluster, one of which is enhanced in $s$-process 
elements and Fe relative to the other \citep[][see also \citet{RMS11}]{AM09, AM11}.   With M$_{V}$ = --8.5
\citep{H96}\footnote{Data values used are those from the latest version of the \citet{H96} catalogue available 
at http://physwww.physics.mcmaster.ca/$\sim$harris/mwgc.dat.},
 M22 is not especially luminous (as M54 and $\omega$~Cen are) and it is kinematically a 
typical inner-halo globular cluster \citep{D99}.  It is not known to be associated with any dwarf galaxy remnant or stellar stream.

Three other Galactic globular clusters also show significant (i.e., $\sigma_{int}$[Fe/H] $\ga$ 0.05 dex)
intrinsic [Fe/H] ranges.  
These are the metal-rich bulge cluster 
Terzan 5 \citep{FDM09} and the intermediate metallicity clusters NGC~1851 \citep{Ca11} and NGC~3201
\citep{Sm13}.  Terzan 5 shows
two distinct stellar populations that differ by $\sim$0.5 dex in [Fe/H] and which also have different 
$\alpha$-element to iron abundance ratios, suggesting a complex enrichment history \citep{OR11}.
NGC~1851, on the other hand, shows a relatively small intrinsic iron abundance dispersion characterised 
by $\sigma_{int}$([Fe/H]) $\approx$ 
0.07 dex \citep{Ca11}.  There is, however, a tendency for the more iron-rich stars to have also higher
abundances of the $s$-process element Ba \citep{Ca11}, a situation reminiscent of the two 
(Fe, $s$-process) groups
in M22.  NGC~1851 has a number of other abundance and photometric peculiarities
\citep[e.g.,][]{AMi08,DY09,Ca11,Ca12}.  Potential scenarios to explain the observations include the
possibility that the current cluster is the product of the merger of two separate  globular clusters 
within a (now
disrupted) parent dwarf galaxy \citep[][and the references therein]{Ca11}.  The presence of an
extensive stellar halo around NGC~1851 \citep{Oz09, CBB12} may be a remnant of the disrupted
dwarf, although the recent results of \citet{So12} show that the situation is complex, with the stellar
halo apparently containing more than one component.  As regards NGC~3201, which has a highly 
retrograde orbit and a luminosity M$_{V}$ = --7.45 \citep{H96}, somewhat fainter than that of NGC~1851
(M$_{V}$ = --8.3), the high dispersion spectroscopic analysis of \citet{Sm13} for 24 member red giants
revealed a total [Fe/H] abundance range of $\sim$0.4 dex with $\sigma_{int}$([Fe/H]) $\approx$ 0.1 dex.  
However, more recently, \citet{MGV13} did not find any strong evidence for a substantial abundance range 
in this cluster from their high dispersion study of 8 NGC~3201 red giants: the dispersion in the [Fe/H] values 
was only 0.04 dex and the range in the observed abundances was 0.12 dex. Similarly, \citet{CB09} and 
\citet{Sa12} also did not find any significant evidence for an abundance range in this cluster.

In the above discussion we have left aside clusters such as M15, which shows an intrinsic spread 
in neutron-capture elements, particularly the $r$-process element Eu, but not in [Fe/H] 
\citep[][and references therein]{SKS11}.  Indeed \citet{DY13} propose that the Galactic globular clusters
can be grouped into three classes: (i) those that exhibit abundance variations in the light elements only;
(ii) those which in addition to the light element variations show a range in the abundances of
neutron-capture elements; and, (iii) those which also possess an intrinsic dispersion in the Fe-peak element
abundances.
The one exception to this scheme is the luminous extreme outer halo 
globular cluster NGC~2419, which appears to be a very different stellar system as regards its elemental
abundances.  \citet[][and the references therein]{MBl12} show that the red giants in this cluster 
have very similar [Fe/H], [Ca/Fe] and [Ti/Fe] abundance and abundance ratios while at the same time
possessing large anti-correlated variations in Mg and K abundances.  The origin of these abundance
anomalies is not easily explained \citep{MBl12}.

In a recent paper \citet{Sa12} reported that the relatively unstudied cluster NGC~5824 could be a
further Galactic globular cluster with an intrinsic [Fe/H] abundance range.  This object is a 
metal-poor globular cluster that lies 32 kpc from the Sun and 26 kpc the Galactic Center in the 
outer halo \citep{H96}.  With M$_{V}$ = --8.9 it is a relatively luminous system: for the globular clusters
further from the Galactic Center than M54 (i.e., R$_{GC}$ $\ga$ 20 kpc), only NGC~2419 is more 
luminous \citep{H96}.  NGC~5824 is also highly centrally concentrated ($c$ = 2.0) with a 
small core-radius of 0.6 pc \citep{H96}.    
The cluster is of further interest because of its possible association with the Cetus Polar Stream
\citep{HN09}.  The Cetus Polar Stream is low luminosity metal-poor
stellar stream in the south Galactic gap characterised by blue horizontal branch stars \citep{HN09, SK12}.
\citet{HN09} did not observe the stream in the vicinity of NGC~5824, which is in fact in the opposite 
Galactic hemisphere, but noted that the cluster has a location, distance, and radial velocity 
consistent with the predictions of the best-fit orbit to their Cetus Polar Stream data \citep{HN09}.  The
metallicity is also consistent with that of the stream stars \citep{HN09}.

NGC~5824 was included in the \citet{Sa12} study because the only previous metallicity estimate was based on an integrated light spectrum.  Intermediate-resolution spectra at the Ca II triplet were obtained
for 17 red giant branch members.  The line strength data showed a dispersion that was notably larger
than that expected from the measurement errors alone, leading to the suggestion of the presence of an
internal abundance spread characterised by $\sigma_{int}$([Fe/H]) $\approx$ 0.12 dex \citep{Sa12}.
As noted by \citet{Sa12}, NGC~5824 was the only cluster besides M22 and M54 in the sample of
20 program and 8 `standard' clusters
studied to show a spread in line-strengths in excess of that expected from the errors.  

Given the relatively small numbers of Galactic globular clusters with intrinsic [Fe/H] distributions, it
is important to verify the \citet{Sa12} results.  We present here the outcome of an 
intermediate-resolution spectroscopic study of a much
larger sample of NGC~5824 red giants than that presented in \citet{Sa12}.
The observations and data analysis techniques are discussed in the following section, while in \S 3
we pay careful attention to the cluster membership status for the stars in the observed sample.    The 
results for the final set of cluster red giant probable members are presented in \S 4 and are discussed
in a broader context in the final section.

\section{Observations and Reductions}

\subsection{Sample Selection}

Observing time to follow-up the possibility of an intrinsic [Fe/H] spread in NGC~5824 was allocated 
for intermediate-resolution spectroscopy of candidate red giant members on both the 
VLT with the FORS2 multi-object spectrograph, and on the Gemini-S telescope with the 
GMOS-S multi-object spectrograph.  The targets for the 
GMOS-S observations were chosen from the photometry derived from the CCD images of the 
cluster obtained as pre-imaging for the FORS2 observations discussed in \citet{Sa12}.
In this case the images were centered on the cluster.   For the new FORS2 observations, a further set of 
pre-images (30 s exposures in $V$ and $I$) was obtained that covers a larger area around the cluster.  
In both cases the list of potential targets was restricted to those stars that lie 
relatively close to the cluster red giant branch (RGB) in the 
colour-magnitude diagrams derived from the imaging data.  The selected stars were also chosen to
cover a magnitude range from the 
vicinity of the RGB tip to approximately 0.5 mag brighter than the horizontal branch magnitude.  

The photometry derived from the new FORS2 pre-imaging, which was obtained over a number 
of nights and under different seeing conditions, was 
subsequently brought on to a single system using stars in the field-to-field overlap regions, and
calibrated to the standard $V$, $I$ system using stars whose standard magnitudes are available
in the photometric standard star fields database maintained by Stetson\footnote{http://www3.cadc-ccda.hia-iha.nrc-cnrc.gc.ca/community/STETSON/standards/}.  The stars observed with GMOS-S are
included in this calibrated photometry set.

\begin{figure}
\centering
\includegraphics[angle=-90.,width=0.46\textwidth]{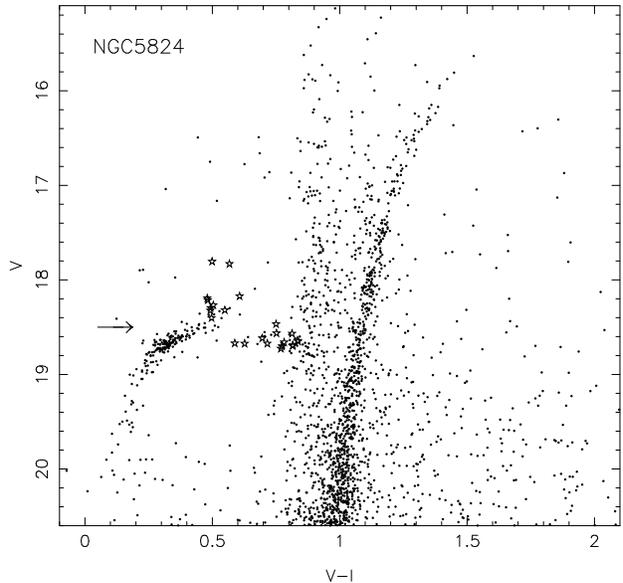}
\caption{Color-magnitude diagram for NGC 5824 from the FORS2 pre-imaging data.  Only stars outside 
50$\arcsec$ from the cluster center are shown.  The maximum radial distance is 8.1$\arcmin$.
The horizontal arrow indicates the adopted value
of $V(HB)$ for the cluster.  Known RR~Lyrae variables in the cluster which occur in the FORS2 
photometry list are plotted as open-star symbols. \label{cmd_fig}}
\end{figure}

In Fig.\ \ref{cmd_fig} we show the final calibrated colour-magnitude diagram (CMD) for NGC~5824 
from the FORS2 pre-imaging.  The conventional definition of $V(HB)$, the mean magnitude of the 
horizontal 
branch, for clusters like NGC~5824 that have a strong blue horizontal branch (HB) population, is the 
mean magnitude of the HB stars at the blue edge of the RR~Lyrae instability gap.  The blue edge lies at
$(B-V)_{0}$ $\approx$ 0.22 in metal-poor globular clusters \citep[e.g.,][]{AS06} which, allowing for
the NGC~5824 reddening of $E(B-V)$ = 0.13 \citep{H96}, corresponds to $(V-I)$ $\approx$ 0.47 mag.
Our CMD then suggests for NGC~5824 $V(HB)$ = 18.50 with an uncertainty of $\pm$0.03 mag.  
This value is in good
agreement with the value $V(HB)$ = 18.45 tabulated by \citet{H96}, which is derived from the CMD study
of \citet{Br96}.  It is also consistent with NGC~5824 CMD derived from $HST$ WFPC2 photometry in
\cite{GP02}.  Also shown in the figure is the photometry from the pre-imaging data for 24 of the 27 NGC~5824
variable stars (all candidate RR~Lyrae variables) listed in the on-line version of the Globular Cluster Variable
Star Catalogue\footnote{http://www.astro.utoronto.ca/$\sim$cclement/read.html} \citep{Cl01}  that we were 
able to unambiguously identify via a position match.  It is 
likely that the search for variables in this cluster is not complete.

\subsection{Spectroscopic Observations}

\subsubsection{VLT}

The spectroscopic observations of the candidate NGC~5824 red giants defined from the pre-imaging data
were carried out in Service Mode with the FORS2 instrument \citep{Ap98}
at the Cassegrain focus of VLT1/UT1-Antu under ESO program 087.D-0465.  The instrumental setup 
employed was identical to that used in \citet{Sa12}: MOS-mode with 1$\arcsec$ slit widths and with
the 1028z+29 
grism plus the OG590+32 order-blocking filter providing a maximum spectral coverage of 
$\sim$$\lambda$7700--9500\AA\/ which includes the Ca II triplet feature.  The MOS-mode
uses 19 moveable slitlets of 20$\arcsec$ length that are placed on the target stars.  A total of 14
MOS-configurations were observed over seven nights in the period 9 May 2011 to 27 June 2011.  
Each observation consisted of a pair of 480 s integrations.  One of the MOS configurations observed was 
deliberately chosen to be identical to that observed in \citet{Sa12}.  Bias frame, arc lamp and flat-field 
exposures were also obtained in the morning following each observing night as part of the standard 
calibration procedures.

\subsubsection{Gemini South}

Observations of candidate NGC~5824 red giants were also carried out in Queue-Scheduled mode on the 
\mbox{Gemini-S} telescope under program GS-2011A-Q-47 using the GMOS-S multi-object spectrograph.  
The spectrograph was configured with the R831 grating and a RG610 filter and set to a central 
wavelength of 
8600\AA.  Four masks were observed on 6 May 2011 with a fifth observed on 13 May 2011.  For both 
observations the seeing was $\sim$1$\arcsec$ with relatively clear skies, consistent with the requested 
conditions.  Each mask used a slit width of 1$\arcsec$ and a slit length 10$\arcsec$ to allow adequate 
subtraction of the bright night-sky emission line features.
A total of 76 candidate red giant branch stars (plus 3 acquisition stars per mask) were allocated 
across the 5 masks.  One star was common to four masks, two were common to three masks and one 
common to two masks.  The
sample also included NGC~5824 red giants originally observed in \citet{Sa12} to ensure line strength
measures from the GMOS-S spectra could be placed on the same system as that of the FORS2 spectra.

The observation of each mask consisted of a 1200 s integration at the central wavelength of 8600\AA\/ 
followed by a second 1200 s integration at a central wavelength of 8550\AA, with each observation 
followed or preceded by a flat field integration at the corresponding central wavelength.  The
observations with different central wavelengths allow for the possibility of a Ca II triplet feature falling 
on the gaps between the three GMOS-S CCD detectors.   Arc lamp exposures for each mask were 
obtained as part of the routine baseline calibration procedures.
 
\subsection{Data Reduction}

\subsubsection{FORS2} 

The FORS2 spectral data were reduced using version 4.6.3 of the FORS2 pipeline \citep{IL08} in an 
identical manner
to that described in \citet{Sa12}.  The two exposures per mask were average-combined after the 
pipeline processing and the final reduced spectra have a resolution of $\sim$3.5\AA\/ at $\lambda$8600\AA\/
and a scale of 0.82 \AA\/ per pixel.  A total of 233 useable spectra of 172 stars resulted.

\begin{figure}
\centering
\includegraphics[angle=-90.,width=0.46\textwidth]{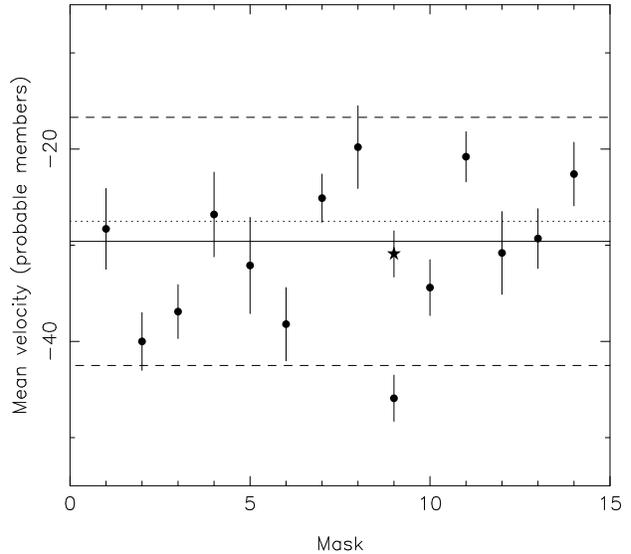}
\caption{The mean heliocentric velocity for NGC~5824 probable members is shown as a filled circle 
for each of the fourteen masks observed with FORS2.  The error bar associated with each point is the 
standard deviation of the mean.  The dotted line is the NGC 5824 cluster velocity given in \citet{H96}, 
while the solid line is the average of the mask mean velocities, excluding mask 9.  The dashed lines are 
$\pm2\sigma$ about this average value.  The star symbol shows the location of the mean velocity for
mask 9 after an adjustment of +15 km s$^{-1}$. \label{FORS2_rvfig}}
\end{figure}

The radial velocities for the stars in each of the 14 FORS2 masks were measured using the three lines 
($\lambda\lambda$ 8498, 8542 and 8662\AA) of the Ca II triplet and the IRAF package {\sc rvidlines}
following an identical process to that adopted in \citet{Sa12}.  In Fig.\ \ref{FORS2_rvfig} we show the mean 
heliocentric velocity for the probable NGC~5824 members in each mask.  In general there is reasonable
agreement between these mean mask velocities and the \citet{H96} catalogue velocity 
\mbox{(--27.5} $\pm$ 1.5
km s$^{-1}$) for the cluster.  The one exception is mask 9 
where the mean velocity of the probable members
is $\sim$15 km s$^{-1}$ lower than the mean velocity for the other 13 masks.  Consequently, we have 
adjusted the velocities for the stars observed in this mask by +15 km s$^{-1}$ to place them in accord with
the other FORS2 velocities.  We note in particular that unlike the earlier NGC~5824 velocities discussed in 
\citet{Sa12}, we see no disagreement between the FORS2 velocities derived here and the \citet{H96} 
catalogue velocity for the cluster.

\subsubsection{GMOS-S} \label{sect_2.3}

\begin{figure}
\centering
\includegraphics[angle=-90.,width=0.46\textwidth]{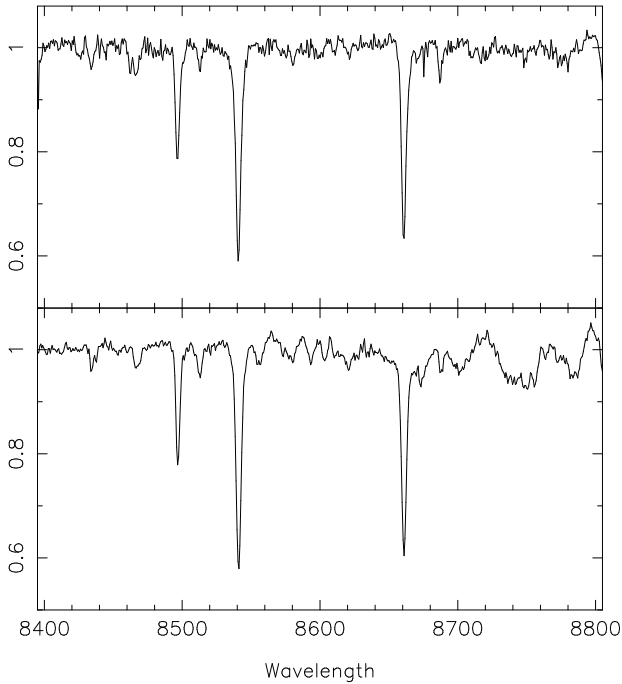}
\caption{Examples of fully reduced GMOS-S spectra from the observations of Mask 1 at a central
wavelength of 8600\AA.  The upper panel shows the spectrum of star 42008180 ($V$=16.51) in which
the fringing has been substantially removed, while the lower panel shows the spectrum of star
42012704 ($V$=16.19) for which significant residual fringing remains.  \label{GMOS_sp}}
\end{figure}

The CCD detectors in GMOS-S at the time of the observations fringe badly at wavelengths beyond 
$\sim$7000\AA.  It is the ability
to deal with this large amplitude fringing that ultimately sets the signal-to-noise of the reduced spectra 
rather than the actual signal.  The data frames for each mask were reduced following example IRAF
Gemini package scripts, modified during the course of the process to achieve as best a reduction in the
fringe amplitude as possible.  In this process the frames with the different central wavelengths, which 
have the Ca II triplet features at different locations on the detectors and which are thus subject to different 
fringing, were kept separate.  The final wavelength-calibrated, sky-subtracted spectra have a resolution
of $\sim$3.5\AA\/ and a (binned) pixel-scale of 0.68 \AA\/ per pixel.  Examples of GMOS-S spectra where the
fringing compensation has worked well, and where it has not, are shown in Fig.\ \ref{GMOS_sp}.

As for the FORS2 spectra, initial radial velocity estimates for each of the stars observed with GMOS-S 
were determined by measuring with {\sc rvidlines} the Ca II triplet line centers on the 
reduced spectra.  The velocities from the 
$\lambda$8600\AA\/ and $\lambda$8550\AA\/ central wavelength observations for each star were then  
averaged.  Since the arc lamp exposures for each mask and central wavelength setting
were carried out during 
daylight hours (and in one case on a different day) the zero point of the velocity scale for each mask is 
not necessarily well determined.  Consequently, we have used the stars in common between the set of
five masks to determine relative mean velocities for each mask, and then assumed that after correction
for the mean mask-to-mask offsets, the mean velocity of the likely members on each mask 
corresponds to the 
cluster velocity given in the \citet{H96} catalogue.  In detail, the velocities were placed on a uniform
system by adjusting those for mask 1 by --15  km s$^{-1}$ and those for mask 5 by +7 km s$^{-1}$ while 
leaving those for masks 2, 3 and 4 unaltered.  A final overall correction of +37 km s$^{-1}$ (equivalent
to a shift of 1.5 pixels) was then applied to place all the velocities on a system that reproduces the 
\citet{H96} catalogue velocity for the cluster.  
 
Two independent checks of this process are possible.  First, there are six stars observed with GMOS-S
that were also observed in \citet{Sa12}.  For these stars the mean velocity difference between the GMOS-S 
and \citet{Sa12} observations is --13 km s$^{-1}$ with a standard deviation of 10.2 km s$^{-1}$.  
Given that \citet{Sa12}
suggest velocity errors of $\sim$5--6 km s$^{-1}$, the standard deviation indicates that the GMOS-S
velocity errors are of similar size.  The offset, not surprisingly given the way the GMOS-S velocities have
been adjusted, accounts for the difference between the mean velocity found for NGC 5824 in \citet{Sa12}
and that in the \citet{H96} catalogue.  The second check is provided by a comparison of the velocities
for the 25 stars that have a radial velocity determination from both the GMOS-S observations and from
the current
FORS2 observations.  These stars have a reassuringly low mean velocity difference (GMOS-S -- FORS2) of
4.2 km s$^{-1}$ and a standard deviation of 13.8 km s$^{-1}$.  Ascribing the errors equally indicates
individual velocity errors are of order 9--10 km s$^{-1}$, which is sufficient to allow use as a membership
criterion.  This is discussed further in section \ref{rvs}.
  
\subsection{Line Strength Analysis}

\begin{figure}
\centering
\includegraphics[angle=-90.,width=0.46\textwidth]{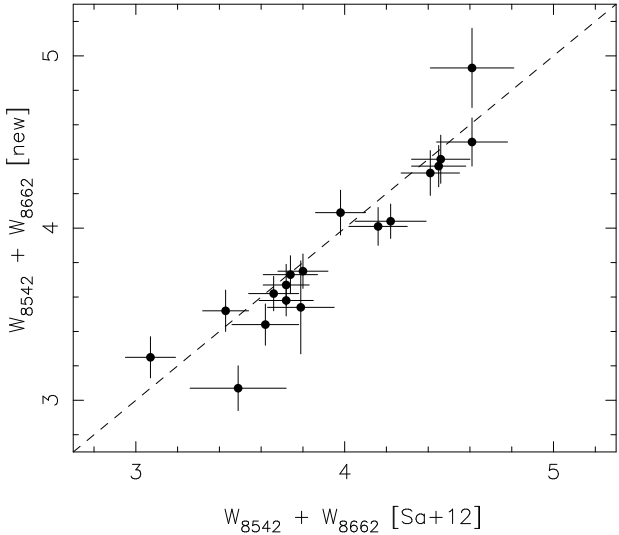}
\caption{The sum of the equivalent widths of the $\lambda$8542\AA\/ and $\lambda$8662\AA\/ lines of the
Ca II triplet ($\Sigma$W) as measured from the new FORS2 observations are plotted 
against the original values from
\citet{Sa12} for the 18 (17 NGC~5824 members, one non-member) stars in common.  
The dashed line indicates 1:1 correspondence and the error bars 
are those returned by the line strength measurement code. \label{fors_old_new}}
\end{figure}

The strengths of the $\lambda$8542\AA\/ and $\lambda$8662\AA\/ lines of the Ca II triplet were measured 
on both the GMOS-S and the FORS2 spectra using identical techniques to those described in \citet{Sa12}.  
In Fig.\ \ref{fors_old_new} we show a comparison of the summed line strengths
($\Sigma$W) for the NGC~5824 stars observed in \citet{Sa12} with those derived from our new FORS2
observation that employed the identical FORS2 slit configuration.   The agreement between the two data sets is excellent with the mean difference between
the values of $\Sigma$W being 0.06 \AA, in the sense \citet{Sa12} minus the new determinations.  
The standard deviation of the differences,
0.17 \AA, is consistent with that expected from the measurement errors.  We can therefore be confident
that line strength measurements made from the new FORS2 spectra are consistent with those in
\citet{Sa12}.  In particular, we can directly apply the abundance calibration developed in \citet{Sa12} 
to the new FORS2 data.

\begin{figure}
\centering
\includegraphics[angle=-90.,width=0.46\textwidth]{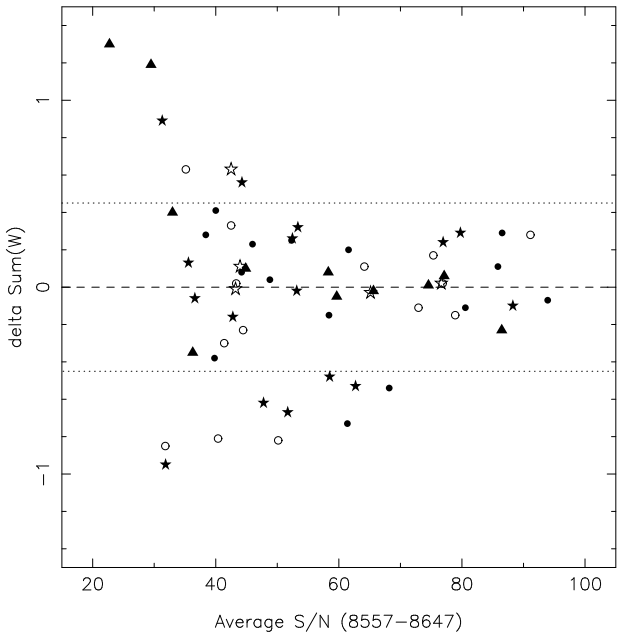}
\caption{The differences in summed equivalent widths of the $\lambda$8542\AA\/ and 
$\lambda$8662\AA\/ Ca II triplet lines between measurements made on GMOS-S spectra centered at 
$\lambda$8600\AA, and those made on spectra centered at $\lambda$8550\AA, are plotted against
the average signal-to-noise (S/N) for each pair of spectra.  The S/N values are computed over the 
wavelength interval
$\lambda$8557--8647\AA.  Stars from Mask 1 are shown as filled circles, Mask 2 as filled stars,
Mask 3 as open circles, Mask 4 as open stars, and Mask 5 as filled triangles, respectively.  
The dashed line indicates equality and the dotted lines are $\pm$0.45\AA; stars lying outside these
lines have not been included in the subsequent analysis.  \label{GMOS-8685comp}}
\end{figure}

The situation for the GMOS-S line strength measurements is, unfortunately, not so clear cut.  
In Fig.\ \ref{GMOS-8685comp} we show the difference between the $\Sigma$W values for
the spectra centered at $\lambda$8600\AA\/ and for the spectra centered at $\lambda$8550\AA,
as a function of the average signal-to-noise (S/N) of the spectra.  We note that since the 
$\lambda$8550\AA\/
observations for Mask 4 were impacted by cloud, the comparison is possible only for the brighter stars
observed with this mask.  Similarly, the comparison is not possible for the small number of stars where
either the $\lambda$8542 or the $\lambda$8662 line fell in an inter-chip gap on one of the individual
spectra.  The S/N of each spectra was calculated from the mean and the 
standard deviation of the counts in the wavelength region $\lambda\lambda$8557--8647\AA\/ (i.e.,
between the two stronger triplet lines).  This wavelength region is relatively free of stellar absorption
lines and the approach takes into account the effect of residual fringing, which 
reduces the S/N below that which would be inferred from the continuum count level alone.

Figure \ref{GMOS-8685comp} shows that for most stars there is reasonable agreement 
between the two measurements
of $\Sigma$W, particularly for the higher S/N spectra.  However, there are also stars where the 
difference is substantial despite apparently good S/N values.  In these cases the residual fringing has
most likely affected either the line profile or the continuum level resulting in an uncertain line strength 
measurement.  Consequently, we have excluded from the subsequent analysis all stars observed
with GMOS-S for which the absolute value of the difference between the two $\Sigma$W values
exceeds 0.45\AA: within this limit the uncertainty in the average of the two determinations is nominally less
than 0.23\AA, which is the largest $\Sigma$W error given in \citet{Sa12} for their NGC~5824 members.
This cut reduces the GMOS-S sample to 45 spectra of 37 stars but given the science aim of the program, 
the increased confidence in the reliability of the line strength measures outweighs the reduction in sample 
size.  For the 45 spectra pairs with $\Delta$($\Sigma$W) $\leq$ 0.45\AA, the mean difference is 
0.05\AA\/ with a standard deviation of 0.20\AA.  The lack of any significant difference between the 
$\Sigma$W measures from the two sets of spectra indicates that they can be meaningfully averaged, and 
all subsequent use of the GMOS-S $\Sigma$W values refers to the averaged value.

\begin{figure}
\centering
\includegraphics[angle=-90.,width=0.46\textwidth]{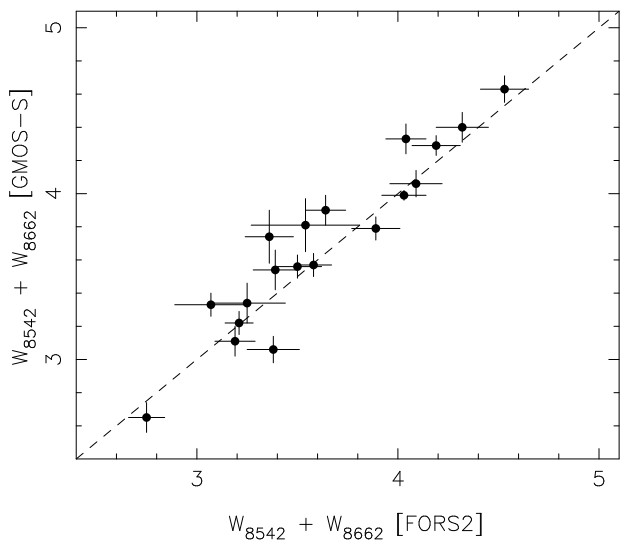}
\caption{The corrected line strengths from the GMOS-S spectra are plotted against the line strengths
from the current FORS2 spectra for the 19 stars in common between the two data sets.  The dashed
line shows 1:1 correspondence and the error bars are those returned by the line strength
measurement code.  \label{GMOS_vs_FORSnew}}
\end{figure}

We can now compare the GMOS-S $\Sigma$W values with those from the FORS2 data.
There are 5 stars in the remaining GMOS-S sample that are also in \citet{Sa12}.  After applying a line
strength correction factor of 1.046 to the GMOS-S values \citep[see][]{Sa12}, the five stars have a 
mean difference in $\Sigma$W,
in the sense \citet{Sa12} minus the GMOS-S determinations, of --0.01\AA. 
The standard deviation of the differences, 0.10\AA, is again consistent with that expected from the 
measurement errors.  This suggests that the corrected GMOS-S values are on the system defined in 
\citet{Sa12}, but a more meaningful comparison is provided by the 19 stars that are in common 
between the culled GMOS-S sample and the new set of FORS2 observations.  This comparison is 
shown in Fig.\ \ref{GMOS_vs_FORSnew} where it is clear that a 1:1 line is consistent with the 
measurements.  The mean difference in $\Sigma$W, in the sense of the current FORS2 measures
minus the corrected GMOS-S measures, is --0.07\AA\/ with a standard deviation of 0.17\AA.  A formal
least-squares fit to these data has a slope of 1.03 $\pm$ 0.09 \AA\/ per \AA, an intercept of -0.03 $\pm$
0.32 \AA, and an $rms$ of 0.17 \AA.  We conclude therefore that the corrected GMOS-S $\Sigma$W 
measurements are on the same system as the current FORS2 values and both are consistent with that 
established in \citet{Sa12}.  

\section{Cluster Membership}

In order to fully evaluate the existence of an internal [Fe/H] range in NGC~5824, it is necessary to have a 
sample of  stars for which the membership status is as unambiguous as possible.  For many clusters such
a task is relatively straightforward because the heliocentric radial velocity of the cluster is sufficiently
different from the velocities of non-member field stars that member/non-member classification is readily
achieved.  This is not the case for NGC~5824 as the 
relatively low velocity (--27.5 $\pm$ 1.5 km s$^{-1}$, \citet{H96}) does not offer much discrimination 
against field stars.  Consequently, we have made use of a series of membership criteria in determining 
our final sample of NGC~5824 red giant members.  We now discuss these criteria in turn.

\subsection{Use of the 8807 Mg I Line}

In a recent paper \citet{BS12} showed that the strength of the $\lambda$8806.8\AA\/ Mg~I line, which is 
gravity sensitive and stronger in dwarfs than in giants, could be used in conjunction with the equivalent
widths of the Ca II triplet lines to effectively discriminate metal-poor red giants from foreground dwarfs.
The FORS2 spectra obtained here generally have sufficient S/N that the $\lambda$8807\AA\/  
Mg~I line could 
be straightforwardly measured on the spectra with the same gaussian-fitting code used for the Ca II 
triplet line strength measurements.  Unfortunately this was not the case for the GMOS-S spectra where 
the residual fringing meant that the continuum could not be well enough defined for a sensible 
measurement of the line strength.

\begin{figure}
\centering
\includegraphics[angle=-90.,width=0.46\textwidth]{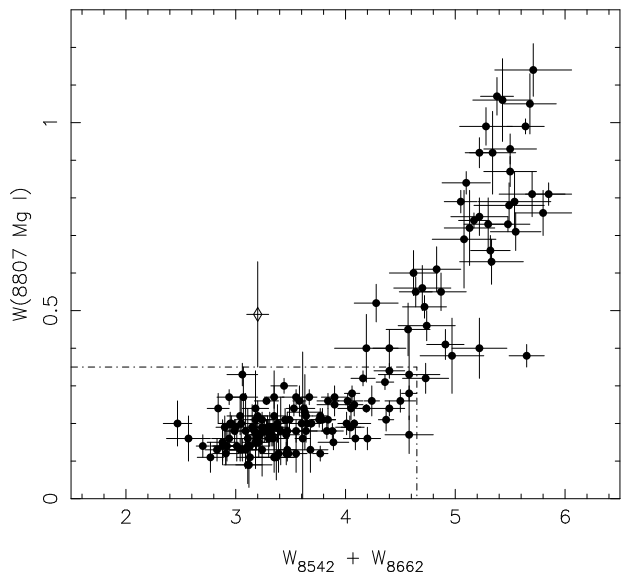}
\caption{The equivalent width of the $\lambda$8807\AA\/ Mg I line, W(8807 Mg I),
is plotted against the sum of the equivalent widths of the $\lambda$8542\AA\/ and $\lambda$8662\AA\/ 
Ca II lines, W$_{8542}$+W$_{8662}$, for 163 stars observed
with FORS2.  The stars inside the region outlined by the dot-dash lines are considered possible cluster
members while the stars outside the region are classified as non-members.  
The star (52005547) with a relatively weak W$_{8542}$+W$_{8662}$ and intermediate but uncertain W(8807 Mg I) 
value, shown by the open diamond symbol, is also 
included in the possible member sample.  \label{mg_vs_cat_final}}
\end{figure}

In Fig.\ \ref{mg_vs_cat_final} we show the strength of the Mg I $\lambda$8807\AA\/ line against $\Sigma$W for
163 of the stars observed with FORS2 (Mg I line strengths could not be measured for 9 FORS2 stars).
As found by \citet{BS12} there are clearly a significant number of stars with relatively large Mg~I 
equivalent widths, which are presumably foreground dwarfs, together with a population showing relatively
weak Mg~I and Ca~II line strengths.  These latter stars are presumably predominantly metal-poor giant 
members of NGC~5824.  After considering the location of the brightest probable cluster red giants in
this diagram, candidate members of NGC~5824 were selected as having Mg~I line strengths weaker
than 0.35\AA\/ and $\Sigma$W values less than 4.65\AA.  The sole exception is star 52005547 for which
the Mg~I line strength measurement is quite uncertain.  This star was retained as a possible member, as
were the 9 FORS2 stars without a Mg~I measurement and the 18 stars in the GMOS-S sample not also in the 
FORS2 sample, leaving a reduced sample of 148 stars.

\subsection{Radial Velocities} \label{rvs}

\begin{figure}
\centering
\includegraphics[angle=-90.,width=0.46\textwidth]{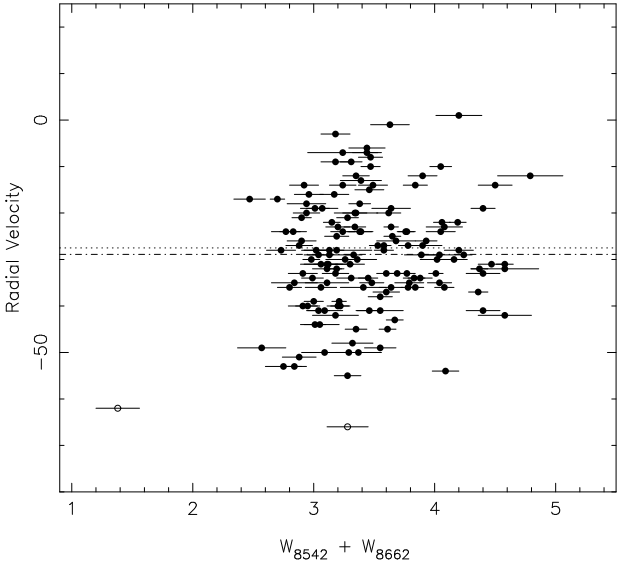}
\caption{Heliocentric radial velocity is plotted against the sum of the equivalent widths of the 
$\lambda$8542\AA\/ and $\lambda$8662\AA\/ Ca II lines, W$_{8542}$+W$_{8662}$, for the all the
stars observed except those excluded from cluster membership via Fig.\ \ref{mg_vs_cat_final}.  Probable
cluster members are shown as filled symbols while the open symbols show stars considered probable
non-members on the basis of their velocity.  The dot-dash line shows the mean velocity of the 145 
probable members while the dotted line shows the \citet{H96} catalogue velocity.  
\label{rv_vs_cat_final}}
\end{figure}

Inspection of the radial velocities of the 148 stars that passed the Mg~I selection criteria reveals one 
obvious outlier -- star 62000110 with a velocity of +288 km s$^{-1}$.  The radial velocities of the 
remaining 147 stars
are plotted against $\Sigma$W in Fig.\ \ref{rv_vs_cat_final}.  Consideration
of this figure suggests two further non-members: stars 12002438 and 41001292 which are plotted as open 
symbols in Fig.\ \ref{rv_vs_cat_final}.  The velocities of these
two stars are notably lower than those of the rest of the sample.   Excluding these two stars the 145 remaining 
candidate members (shown as filled symbols) have mean velocity of --28.9 km s$^{-1}$ and a standard
deviation $\sigma_{v}$ of 11.7 km~s$^{-1}$.  Both 12002438 and 41001292 then lie further than 3$\sigma_{v}$ 
from the mean.  The value of $\sigma_{v}$ is consistent with the combination of the individual velocity errors 
(9--10 km s$^{-1}$, see \S \ref{sect_2.3}) and a radial decrease in the intrinsic velocity of dispersion of 
NGC~5824 stars from the central value of 11 $\pm$ 2 km s$^{-1}$ \citep{DMM97}.  The mean velocity 
for the sample of --28.9 $\pm$ 1.0 km~s$^{-1}$ is fully consistent with the value of --27.5 $\pm$ 1.5 km~s$^{-1}$ 
listed in the \citet{H96} catalogue.

\subsection{Color-Magnitude Diagram}

\begin{figure}
\centering
\includegraphics[angle=-90.,width=0.46\textwidth]{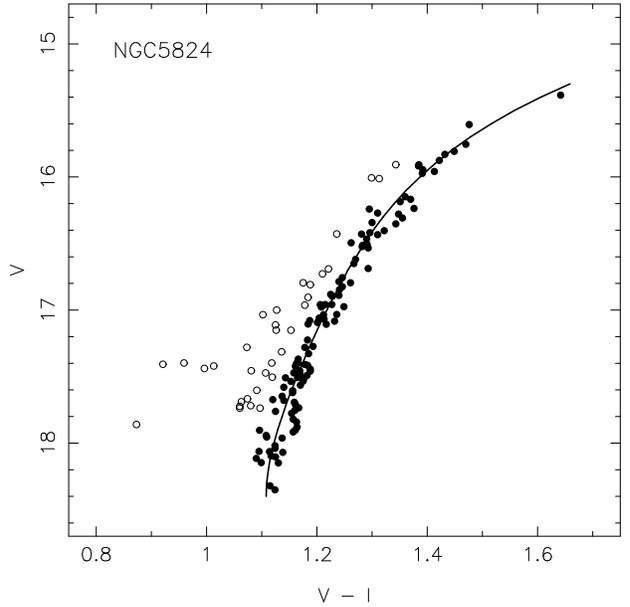}
\caption{Color-magnitude diagram for the 145 NGC 5824 probable cluster members that satisfy 
the (Mg~I, Ca~II)
line strength and radial velocity criteria.  Filled circles are used for candidate members on the RGB
and open circles for candidate AGB members.  The solid line is a fourth order polynominal fit to the RGB 
stars.  
\label{cmd_fig2}}
\end{figure}

In Fig.\ \ref{cmd_fig2} we show the CMD for the  candidate NGC~5824 members that have
passed the (Mg~I, $\Sigma$W) line strength (where a Mg~I line strength has been measured)
and radial velocity selection criteria.  
We note first that, as is immediately apparent from the figure, there exist a sizeable number of stars that lie 
away from the cluster RGB in the CMD\@.   Many of these were included in the FORS2 
observations in order to ensure a broad colour selection to prevent any bias in the [Fe/H]
determinations.  A number of these stars are clearly
asymptotic giant branch (AGB) members of the cluster.  Because of their lower mass, 
AGB stars have lower gravities than RGB stars and at the same colour, or
effective temperature, AGB stars are also brighter than RGB stars.  This offset can be significant at
luminosities fainter than the vicinity of the RGB tip.   Consequently, it is clear that AGB stars will occupy 
a different location in the customary line strength analysis plot \citep[e.g.,][]{Sa12}, which uses $V-V_{HB}$
as the abscissa, even if the actual abundance of the AGB stars is the same as that for the RGB stars.  
We have therefore visually identified 33 AGB star candidate members in the CMD of Fig.\ \ref{cmd_fig2},  and
these stars are shown as open symbols.  Shown also on the figure is a fit of a fourth order polynomial to the
RGB stars that defines the locus of the mean RGB in the CMD\@.
The line-strength analysis is then restricted to the 112 RGB candidate members.

\subsection{Line Strength Diagram}

\begin{figure}
\centering
\includegraphics[angle=-90.,width=0.46\textwidth]{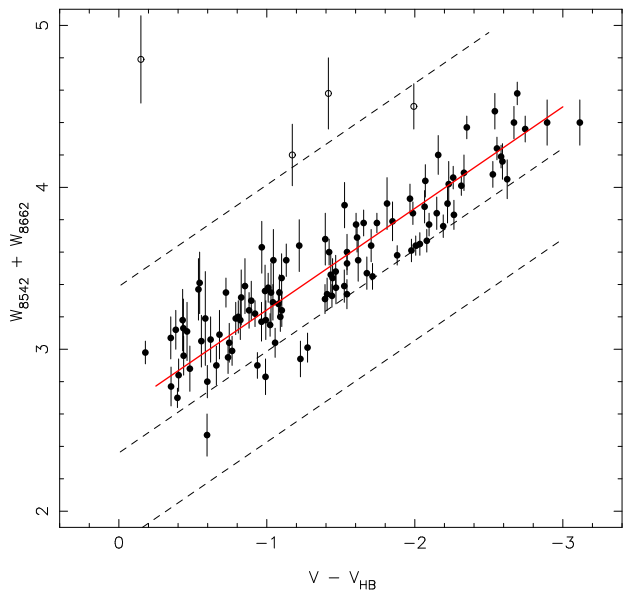}
\caption{Sum of the equivalent widths of the $\lambda$8542\AA\/ and $\lambda$8662\AA\/ lines of the
Ca II triplet is plotted against $V-V_{HB}$ for the 112 candidate RGB members of NGC~5824.  Four
stars that have line strengths substantially stronger than the other stars are plotted as open-symbols.
These stars are probably not cluster members.   The dashed lines show the relation between line
strength and $V-V_{HB}$ for abundances of [Fe/H] = --2.3, --2.1 and --1.7 dex using the calibration of
\citet{Sa12}. The solid red line is the fit to the filled symbol points and corresponds to a mean abundance
for NGC~5824 of [Fe/H] = --2.01 dex.
\label{ew_fig}}
\end{figure}

The sum of the equivalent widths of the $\lambda$8542\AA\/ and the $\lambda$8662\AA\/ Ca II triplet lines,
$\Sigma$W, is plotted against magnitude difference from the horizontal branch, 
$V-V_{HB}$, for the 112 candidate RGB members of NGC~5824 in Fig.\ \ref{ew_fig}.  
Shown also on the plot 
as dashed lines are the (line strength, magnitude) relations for abundances [Fe/H] = --2.3, --2.0 and --1.7 
dex  as implied by the abundance calibration derived in \citet{Sa12}.  Four stars stand out as having
line strengths notably stronger than the rest of the sample.  As inferred from Fig.\ \ref{ew_fig} these stars 
have abundances of [Fe/H] $\approx$ --1.7 dex or larger and should therefore lie to the red of the mean RGB in
Fig.\ \ref{cmd_fig2}.  However, this is not the case --- all four lie within the dispersion about the mean 
RGB\@.  Consequently, we shall assume that these four stars (41001144, 42006240, 42011950 
and 61002588) are not cluster members and not consider them further.  We note, however, that the star 
42006240 is star 2\_32429 in the \citet{Sa12} study and was considered a cluster member in that work.
The final sample is then 108 RGB candidate members.

For completeness we note that the equivalent plot to Fig.\ \ref{ew_fig} for the AGB stars identified in
Fig.\ \ref{cmd_fig2} shows that all the stars, with the sole exception of star 61000920, have line
strengths that are consistent with cluster membership.  Star 61000920 has a much larger 
W$_{8542}$+W$_{8862}$ value than the other AGB stars with similar $V-V_{HB}$ and is thus
unlikely to be a cluster member.  We also note that, as expected, the mean line through the AGB 
data points is $\sim$0.4 \AA\/ weaker at fixed $V-V_{HB}$ compared to the equivalent relation for the
RGB stars.

In Table \ref{Table1} we list the identification number, position, photometry, line strength 
measurements and 
additional information for the entire sample of 190 stars observed at both the VLT and Gemini-S\@.   

\begin{table*}
\begin{minipage}{0.98\textwidth}
\caption{NGC 5824 Stars Observed$^e$ \label{Table1}}
\begin{tabular*}{0.98\textwidth}{ccccccccccccc}
\hline
ID & RA (2000) & Dec (2000) & $V$ & $V-I$ & Rad Vel & $\Sigma(W)^a$ & error & W$_{8807}$ 
 &  error & N$_{F}^{b}$ & N$_{G}^{c}$& Class$^{d}$ \\ 
& & & & & (km s$^{-1}$) & (\AA) & (\AA) & (\AA) & (\AA) &  & &   \\
\hline
11000467 & 226.10370  & --33.06391 & 17.433 & 0.952  & +95 & 4.28 & 0.20 & 0.52 & 0.05  &  1 &  0 & NM \\
11000781 & 226.13540 & --33.05518 & 16.471 & 1.152  & --46 & 5.28 & 0.24 & 0.99 & 0.05 &1 & 0 & NM \\
11001198 & 226.06163 & --33.04313 & 16.352 & 1.343  & --14 & 3.84 & 0.10 & 0.26 & 0.02 & 1 & 0 & MR\\
11001586 & 226.11004 & --33.02972 & 17.897 & 0.953  & --50 & 4.64 & 0.15 & 0.55 & 0.04 & 3 & 0 & NM\\
\hline
\end{tabular*}

\medskip
{\it Notes.}
$^a$W$_{8542}$+W$_{8862}$.
$^b$Number of observations with FORS2 at the VLT.
$^c$Number of observations with GMOS-S at Gemini South.
$^d$MR indicates star is member of the final RGB sample, MA for stars in the final AGB member sample, 
and NM for probable non-member of the cluster.
$^e$This table is available in its entirety in a machine-readable form in the online
journal.   A portion is shown here for guidance regarding its form and content. 
\end{minipage}
\end{table*}

\section{Analysis}

In a process identical to that followed in \citet{Sa12}, an unweighted fit of a line with slope --0.627\AA/mag 
to the final sample of NGC~5824 RGB stars in Fig.\ \ref{ew_fig} yields a $\langle W^{\prime} \rangle$ 
value of
2.62 \AA,  with a standard deviation of the mean of 0.02 \AA.  Using the abundance calibration given in
\citet{Sa12}, this yields a mean abundance for NGC~5824 of [Fe/H] = --2.01 $\pm$ 0.13, where the
uncertainty is dominated by that of the calibration relation.  This abundance is in fact identical to that 
found in \citet{Sa12}.   Again following the same procedures as \citet{Sa12}, the $rms$
dispersion about the fitted line is 0.19 \AA\/ while the mean of the errors returned by the measurement code,
is significantly smaller, at 0.11 \AA\@.  This implies an intrinsic dispersion in the line strengths of 0.15 \AA, or 
$\sigma_{int}$([Fe/H]) = 0.06 dex.  This value is lower than that, 0.12 dex, found in \citet{Sa12} from 
a sample
of 17 stars, as against the 108 employed here.  As noted in \citet{Sa12}, the largest contribution to their
$rms$ dispersion came from two stars: 2\_32429 (42006240), which we now consider a non-member,
and 1\_8575 (42013479) whose line strength is measured here as $\sim$0.4 \AA\/ less than 
the value given 
in \citet{Sa12}.  Removing these two stars from the \citet{Sa12} dataset then results in an intrinsic line 
strength dispersion of 0.16 \AA, which is then completely consistent with the new value found here.  We 
conclude that \citet{Sa12} were correct in reporting that NGC~5824 possesses an intrinsic [Fe/H] range,
although the size of the intrinsic dispersion measured here is considerably smaller.

\begin{figure}
\centering
\includegraphics[angle=-90.,width=0.46\textwidth]{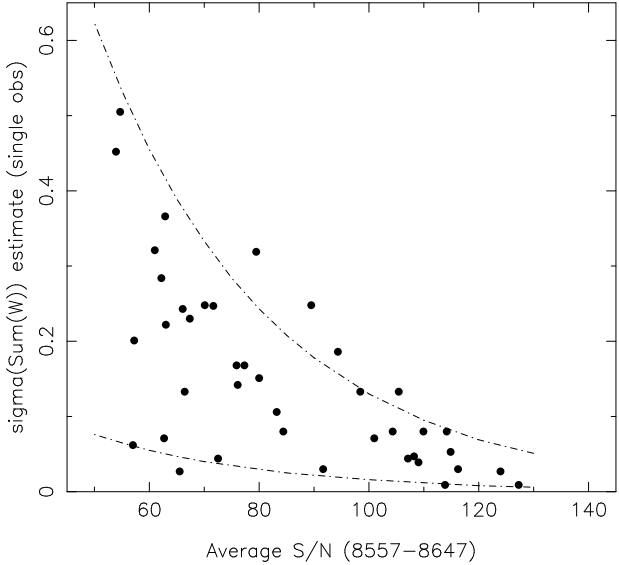}
\caption{Estimates of the error in the summed equivalent width of the $\lambda$8542\AA\/ and 
$\lambda$8662\AA\/ Ca II triplet lines for a single FORS2 observation as a function of the
signal-to-noise ratio in the wavelength region $\lambda\lambda$8557--8647\AA.  The filled circles
are estimates from the stars which have at least two FORS2 observations.  The dot-dash curves are
exponentials that define the adopted upper and lower envelopes to the data points.
\label{fors_err_fig}}
\end{figure}

The above analysis assumes that the line strength errors returned by the measurement code are a 
reasonable estimate of the true errors.  \citet{Sa12} adopted this assumption but
with these new data we can make use of the stars with repeat spectra to gain an independent assessment of 
the line strength measurement errors as a function of signal-to-noise.  
For the FORS2 spectra, there are 40 stars with at
least two observations and we estimate the error in a single FORS2 measurement of $\Sigma$W as follows.  
The error estimate for a single observation is calculated
using the formalism of \citet{K95} in which the error is given by the expression $k$ $\times$ $| \Delta W |$, 
where $| \Delta W |$ is the range of the equivalent width measures for a given star and 
$k$ is a constant that depends on the number of observations, e.g., for two observations $k$ = 0.886.  
In Fig.\ \ref{fors_err_fig} this single observation error estimate is plotted against the average 
signal-to-noise in the wavelength region $\lambda\lambda$8557--8647\AA\/ for the 40 stars with at least
two FORS2 observations.  As noted above, this wavelength
region, which lies between the two stronger Ca~II triplet lines, is relatively free of stellar absorption lines 
and the S/N provides a good measure of the quality of the spectrum.  Shown also on the figure are two 
exponential curves which outline the adopted upper and lower envelopes to the data points.  These have
been derived via a least squares fit to the [S/N, ln (error estimate)] data to determine the exponent, with
the amplitude then increased and decreased to encompass the majority of the points.  For stars with FORS2 
spectra, we then assume that at a given signal-to-noise, the error for an individual star is uniformly distributed 
between the upper and lower limits in the figure.  

Similarly, for the Gemini spectra we use the 
equivalent plot (cf.\ Fig.\ \ref{GMOS-8685comp}) to assess the likely errors in the measurements of the 
summed equivalent widths, and again assume that the errors are uniformly distributed between upper and 
lower boundaries that are a function of the signal-to-noise.  For multiply observed stars within each
sample, the estimated error is reduced by the square root of the number of observations, while for stars
observed with both GMOS-S and FORS2, the uncertainty is that of (unweighted) mean.  Within the FORS2
sample there are 9 stars\footnote{The stars are 31003802, 32004263, 42007205,
42007589, 51001013, 52005984, 52010221, 61000102 and 61000435.} with computed S/N $\leq$ 50, 
beyond the limits of the exponential envelope 
curves in Fig.\ \ref{fors_err_fig}.  We have not considered these stars further leaving a final sample of 99 
NGC~5824 red giants.
We can now simulate the effect of the measurement errors on the summed equivalent 
width determinations, and hence on the abundance distribution.  

\begin{figure}
\centering
\includegraphics[angle=-90.,width=0.46\textwidth]{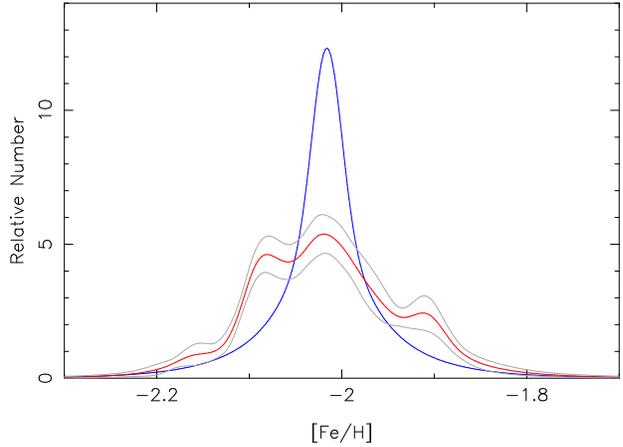}
\caption{Generalized histograms representing abundance distributions derived from the observed
summed equivalent width values.  The red curve is the mean generalized histogram assuming that the
measured summed equivalent widths translate directly to abundance [Fe/H], while the grey curves
show the $\pm$3$\sigma$ deviates about the mean from the sampling of the measurement errors.  
The blue curve is the mean generalized histogram under the assumption that there is no intrinsic 
abundance dispersion among the NGC~5824 red giants. 
\label{abund_dist1}}
\end{figure}

In Fig.\ \ref{abund_dist1} we show the outcome of 1000 trials in which the observed summed
equivalent widths are translated directly to metallicity, with errors assigned as described above.  The 
red line is the mean generalized histogram from these trials while the grey curves
represent the $\pm$3$\sigma$ deviates about the mean.  As a comparison, we assume as a null 
hypothesis that the cluster has a single metallicity equal to the mean determined above and that the 
error for each star is the absolute value of the difference between the observed summed equivalent width 
and the value expected for the mean abundance at the $(V-V_{HB})$ of the star.  The generalised
histogram under this assumption is shown as the blue line.   In both cases the minimum error in
[Fe/H] has been taken as 0.015 dex Áin order to minimise the influence on the overall histograms of 
$\delta$-function-like spikes, which can occur if the error in [Fe/H] becomes unrealistically small.  The
exact value of this limit, within the range 0.010 -- 0.020 dex, does not affect the interpretation of the
histograms.

It is clear from Fig.\ \ref{abund_dist1}, that even allowing for the 
$\pm$3$\sigma$ deviates, the two assumptions provide curves that are significantly different,  confirming that 
NGC~5824 does indeed possess an intrinsic dispersion in [Fe/H].   We can characterise the distribution
shown by the red curve in Fig.\ \ref{abund_dist1} in a number of ways; for example, the FWHM of the
distribution is 0.16 dex.  Alternatively, using the abundance determinations themselves, the inter-quartile 
range (IQR) of the sample is 0.10 dex and the total range in [Fe/H] is $\sim$0.30 dex.  We have also investigated
whether any correlation is present between the deviations in line strength at fixed $V-V_{HB}$ from the 
mean relation in Fig.\ \ref{ew_fig} and deviations in $(V-I)$ colour at fixed $V$ from the mean red giant branch
in Fig.\ \ref{cmd_fig2}.  No convincing correlation is present.  This lack of a correlation, however, is not 
surprising given the small
size of the abundance dispersion and the comparative lack of sensitivity of RGB colour to abundance at low 
abundances.  For example, using the Dartmouth isochrones \citep{DC08}, the difference in $(V-I)$ at
$V$ = 16.0 ($V-V_{HB}$ = --2.5) between an RGB with [Fe/H] = --2.1 and one with [Fe/H] = --1.9 is only 0.033
mag, while at $V$ = 17.5 ($V-V_{HB}$ = --1.0), the difference is even smaller, 0.013 mag.   Intrinsic colour
differences of this order are simply too small to convincingly detect given the errors in the photometry. 

Fig.\ \ref{abund_dist1} also shows that
NGC~5824 abundance distribution is apparently not unimodal -- the red solid line in the figure shows three 
distinct peaks that correspond to abundances of [FeH] $\approx$ --2.09, --2.01 and --1.91 dex, with 
approximately equal numbers in the first  two groups and a smaller population in the third more metal-rich
group.  We have endeavoured to model this observed distribution to investigate its implications.  
The modeling has been carried out by randomly selecting 99 abundances (corresponding to the number 
of stars in the final observed 
RGB sample) from an assumed input abundance distribution, and assigning errors to these abundances in the
same way as for the observed stars allowing computation of a generalised histogram.  
The selection process is then repeated 1000 times and the mean generalised histogram for the model
calculated.  A model is then accepted as a satisfactory fit to the
observations if the model mean generalised histogram is contained entirely within the $\pm3\sigma$ limits
of the observational data histogram shown in Fig.\ \ref{abund_dist1}.  

\begin{figure}
\centering
\includegraphics[angle=-90.,width=0.46\textwidth]{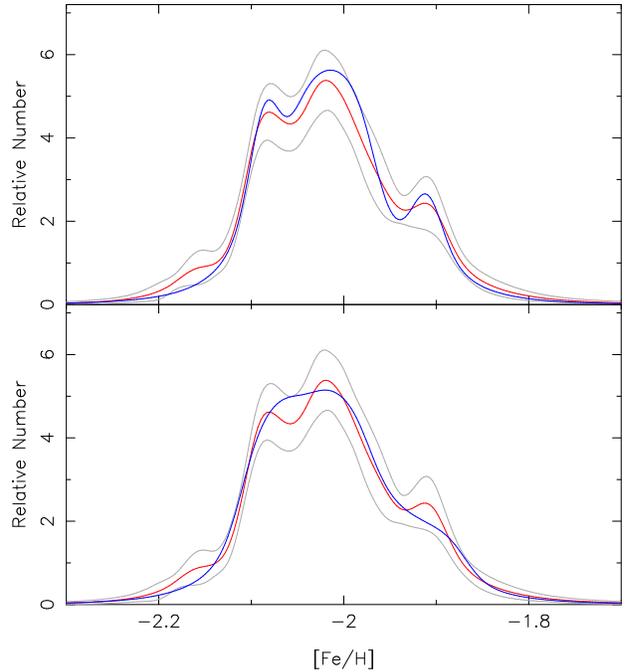}
\caption{As for Fig.\ \ref{abund_dist1} the red curve is the mean generalized histogram for the
observations, generated by assuming the
measured summed equivalent widths translate directly to abundance [Fe/H].  The grey curves
show the $\pm$3$\sigma$ deviates about the mean from the sampling of the measurement errors.  In the
{\it upper panel} the blue curve is the mean generalized histogram resulting from a model which assumes
that there are three separate populations in the cluster: one with [Fe/H] = --2.09 comprising 25\% of the total, 
a second with abundances uniformly distributed between [Fe/H] = --2.06 and --1.97 and 60\% of the total, 
and a third with [Fe/H] = --1.91 and 15\% of the total.  The
{\it lower panel} shows with the blue curve the mean generalised histogram for an alternate model.  In
this model the abundances for 80\% of the population are distributed uniformly between [Fe/H] = --2.11 
and --1.97 dex, while the remaining 20\% have metallicities drawn from a distribution that decreases linearly 
from matching the uniform population relative number at [Fe/H] = --1.97 to zero at [Fe/H] = --1.87 dex. 
\label{abund_dist2}}
\end{figure}

Examples of this process are shown in the panels of Fig.\
\ref{abund_dist2}.  In the upper panel the adopted input abundance distribution has three components: 
the first has a single abundance of [Fe/H] = --2.09 and comprises 25\% of the total, the second has 
abundances uniformly distributed between [Fe/H] = --2.06 and --1.97 and comprises 60\% of the total, and 
the third has a fixed abundance of [Fe/H] = --1.91 and 15\% of the total.  Effectively by construction this
model does an acceptable job of reproducing the three apparent peaks in the observed abundance
distribution.  In this context it is worth noting that with this type of three component model it was not
possible to achieve an acceptable fit in the case where the three components each had discrete abundances.
We must, however, be careful not to over-interpret the outcomes of this heuristic model fitting.  For example,
we show in the lower panel of Fig.\ \ref{abund_dist2} a model which also provides an acceptable fit to the
observed data and which does not contain any discrete components.  For this model 80\% of the population
is assumed to have metallicities distributed uniformly between [Fe/H] = --2.11 and --1.97 dex, while the
remaining 20\% have metallicities drawn from a distribution that decreases linearly from matching the
uniform population relative number at [Fe/H] = --1.97 to zero at [Fe/H] = --1.87 dex. 

There are nevertheless robust conclusions that can be drawn from the modeling process aside from the 
obvious one that an intrinsic abundance spread is required.  The first is that on the metal-poor side the
abundance distribution must rise rapidly with increasing abundance.  It is not possible to determine whether 
a particular abundance cut-off near [Fe/H] $\approx$ --2.1 is required, below which no stars are found, or 
whether there is simply a very rapid increase in numbers above this abundance, but a feature of this nature 
is required.  A sharp rise in the metallicity distribution function on the metal-poor side is remininscent of the
abundance distributions in clusters such as $\omega$~Cen and M22 (see \citet{DC09} and Fig.\  1
in \citet{DM11}).  In this respect these star cluster abundance distributions contrast strongly with those for 
dwarf galaxies
where the abundance distributions show much less steep increases with increasing metallicity on the
metal-poor side of the distribution \citep[see for example, Fig.\ 17 in][]{JN10}.  In contrast to the metal-poor
side, the metal-rich side of the NGC~5824 abundance distribution declines more slowly but the extent 
of the metal-richer population is less than for M22 and much less than for $\omega$~Cen.

The second conclusion that can be drawn from the modeling process is that distributions with strong single
peaks, such as a Gaussian, provide a poor representation of the observations.  Instead, the observations 
suggest input distributions that are more ``flat-topped'', either as discrete components of approximately
similar strength or more uniform distributions such as the model used to generate the lower panel of 
Fig.\ \ref{abund_dist2}.  Unfortunately, given the sample size observed here, it is unlikely that further progress
can be made in characterising the NGC~5824 abundance distribution until a comparable number of 
precise [Fe/H] determinations are made using high dispersion spectroscopy. 

One additional point, however, needs to be discussed.  NGC~2419 is a luminous outer halo globular cluster
whose overall abundance, [Fe/H] = --2.09 \citep[e.g.,][]{MBl12}, is similar to that for NGC~5824.
\citet{MBl12} have shown that while NGC~2419 does show large anti-correlated variations in Mg and K
abundances, it apparently does not possess an intrinsic range in [Fe/H] or [Ca/H] \citep{MBl12}.  This is
despite initial suggestions to the contrary \citep[e.g.,][]{CK10}, which were based on analysis of Ca
triplet line strengths.  \citet{MBl12} instead ascribe the observed  variations in the Ca triplet line strengths
to opacity differences that result from variations in the abundance of the electron-donor element Mg: 
the Mg deficient stars consequently show stronger $\lambda$8542\AA\/ and $\lambda$8662\AA\/ Ca 
lines and vice versa without any true variation in [Ca/H] or [Fe/H]\@.  Therefore, given these results for 
NGC~2419, it is necessary for us to check that our inferred Ca triplet line strength spread is  
due to a real spread in Ca (and Fe) abundances rather than to an opacity effect.  

We have carried out this check by investigating the strength of the $\lambda$8688.6\AA\/
Fe~I line and of the $\lambda$8806.7\AA\/ Mg~I line as a function of $\Sigma$W.  
We considered only the brighter
RGB stars, those with $(V-V_{HB}) \leq -2.0$, and with FORS2 spectra, in order to maximise the 
signal-to-noise of the measurements.  
The Fe~I line was measured via gaussian fitting using the same continuum regions
as for the $\lambda$8662\AA\/ Ca triplet line.  If the NGC~2419 effect was also present in NGC~5824
then we would expect the red giants with stronger Ca triplet lines to have weaker Mg~I lines at fixed
$(V-V_{HB})$, and vice versa, and for there to be no intrinsic scatter in the iron line strengths about 
any mean trend 
with $(V-V_{HB})$.  We find from our measurements that firstly, there is only marginal support for 
any anti-correlation between $\Sigma$W and the strength of the Mg~I  $\lambda$8807\AA\/ line.
After removing a (small) trend with $V-V_{HB}$, we find that the stars with weaker
Ca triplet lines have stronger Mg~I lines at only the $\sim$1$\sigma$ level (5 `weak' stars versus 3 `strong'
stars from a full sample of 22 objects).  Secondly, and more significantly, we find, again after removing a 
small trend with $V-V_{HB}$, that the stars with stronger Ca II lines do indeed have stronger 
$\lambda$8688.6\AA\/ Fe~I lines, at the 2$\sigma$ level, compared to the stars with weaker Ca II lines. 
Consequently, while the outcome is undoubtedly
limited by the signal-to-noise of the spectra and the weakness of the Mg~I and Fe~I lines (typical
measured equivalent widths of 0.24\AA\/ and 0.18\AA, respectively), we see no compelling reason to 
suggest that the observed intrinsic variation in Ca triplet lines among the red giants in NGC~5824 
is anything else other than the result of an intrinsic overall abundance dispersion in this cluster.

\section{Discussion}

\begin{figure}
\centering
\includegraphics[angle=-90.,width=0.46\textwidth]{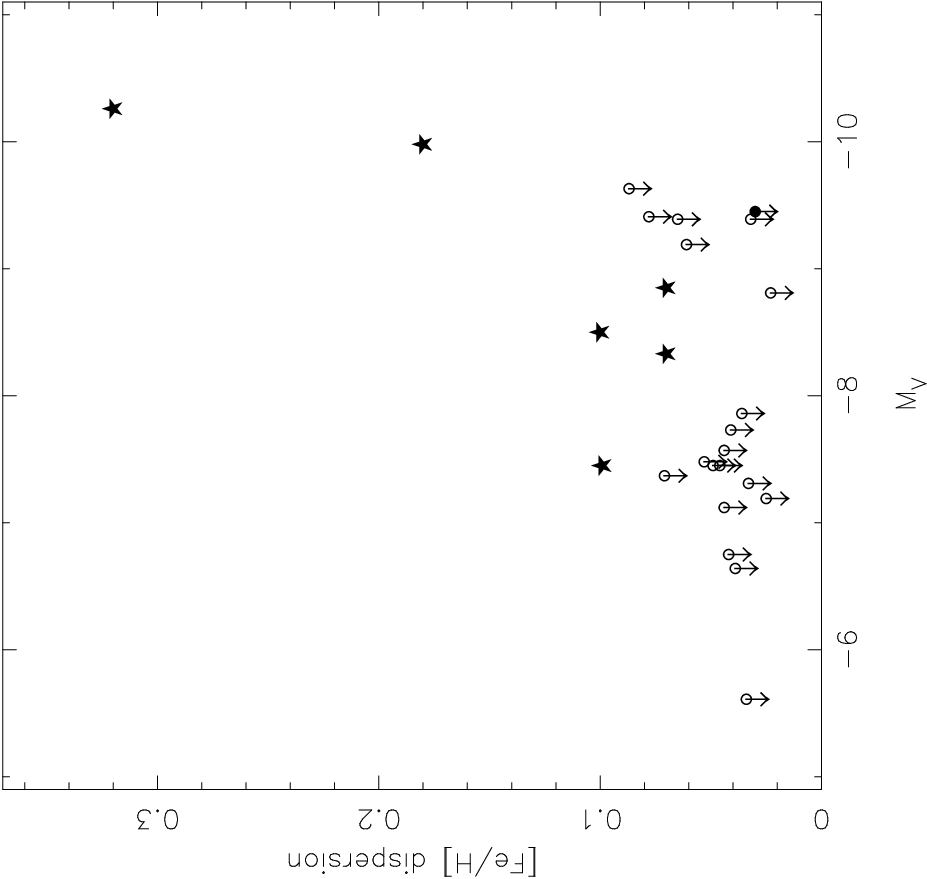}
\caption{Standard deviation in [Fe/H] abundances is plotted against absolute visual magnitude for a number of 
Galactic globular clusters.  The open symbols are upper limits on any intrinsic abundance dispersions and
are taken from \citet{CB09}.  The filled circle is NGC~2419 using the [Fe/H] abundance dispersion limit from
\citet{MBl12}.  The star-symbols are clusters with measured instrinsic [Fe/H] abundance dispersions exceeding
0.05 dex.  In order
of decreasing absolute magnitude the clusters are $\omega$~Cen and M54, from \citet{CB09}, NGC~5824
from this work, M22 \citep{AM11}, NGC~1851 \citep{Ca11} and NGC~3201 \citep{Sm13}.
\label{sprd_vs_Mv_fig}}
\end{figure}

With our results NGC~5824 joins the small list of Galactic globular clusters that possess sizeable intrinsic 
[Fe/H] 
abundance dispersions\footnote{We note that if studies at the exquisite level of precision obtained by 
\citet{DY13} for NGC~6752 red giants were common, it is conceivable that they would reveal many
globular clusters possess intrinsic iron abundance dispersions at the $\sigma_{int}$([Fe/H]) 
$\la$ 0.02--0.03 dex level.}.  
As noted in the Introduction this list also includes $\omega$~Cen, M54, M22, NGC~1851 and 
perhaps NGC~3201
(see \S \ref{Intro} for references); we again choose to exclude the metal-rich Galactic Bulge cluster Terzan~5 
from this list given that the large difference in abundance between the two distinct stellar populations present 
in the cluster \citep{OR11} suggests a different mechanism is involved.  In Fig.\ \ref{sprd_vs_Mv_fig} we show
an updated version of the relation between intrinsic [Fe/H] dispersion, as measured by the standard
deviation of the [Fe/H] distribution\footnote{The IQR of the distributions would be a better measure to employ 
here as it makes no assumption about the form of the distribution, but this statistic is not available for all the
clusters.}, and absolute magnitude for Galactic
globular clusters.  This relation was first discussed in \citet[][see also \citet{CB09}]{CB10a}.  
In the figure upper limits on intrinsic [Fe/H] dispersion values are
shown by open symbols; these data are taken directly from \citet{CB09}.  The filled star-symbols are the
clusters (excluding Ter~5) with identified significant intrinsic [Fe/H] dispersions, which now include NGC~1851,
NGC~3201 and NGC~5824, results that were not available to \citet{CB10a}.  The plotted value for NGC~3201 
comes from the [Fe/H] values tabulated by \citet{Sm13} while that for NGC~1851 comes from \citet{Ca11}.  
The value for NGC~5824 is the one derived here, while for M22  we use the value derived from the [Fe/H] 
abundances listed in \citet{AM11}; \citet{DC09} list a somewhat higher value of $\sigma_{obs}$[Fe/H] = 0.15 dex.  
The other change from the depiction of this figure in \citet{CB10a} is for 
the cluster NGC~2419.  \citet{MBl12} have established that the intrinsic [Fe/H] abundance range in 
NGC~2419 is actually small, in contrast to earlier estimates \citep[see the discussion in][]{MBl12}. 

Although the number of clusters studied is not large, and selection effects are undoubtedly 
important\footnote{For example, \citet{CB09} give an upper limit for the [Fe/H] abundance dispersion in 
NGC~3201 as 0.049 dex and \citet{Sa12} also find no evidence for a [Fe/H] abundance dispersion in the 
cluster.  Both results contrast with the intrinsic dispersion of 0.10 dex found by \citet{Sm13}.  A possible 
explanation for the difference is that the first two studies deliberately selected stars near the mean RGB for 
observation, while the \citet{Sm13} study is less biased in that respect.  See also the discussion in 
\citet{MGV13}.}, 
there does not seem to be any definite correlation between the size of $\sigma_{int}$([Fe/H])  and M$_{V}$ 
in Fig.\ \ref{sprd_vs_Mv_fig}, particularly if the \citet{Sm13} results for NGC~3201 are valid.  Instead, while 
$\omega$~Cen and M54 clearly standout in terms of luminosity and size of abundance
spread, for the other four clusters with intrinsic spreads the size of the intrinsic dispersion does not
clearly change with absolute magnitude. This differs from the suggestion in \citet{CB10a} of an apparently 
statistically significant relation between the size of the intrinsic iron abundance spread and luminosity.  
What is apparent from Fig.\ \ref{sprd_vs_Mv_fig} however, is that the likelihood of an intrinsic abundance dispersion
is larger for more luminous clusters: 5 of the 12 clusters 
in Fig.\ \ref{sprd_vs_Mv_fig} with \mbox{M$_{V}$ $<$ --8} show intrinsic dispersions, while only 1 or none 
(depending on the reality of the NGC~3201 result) of the 14 less luminous clusters show intrinsic dispersions.
Abundance dispersion limits or determinations for a larger sample of clusters are needed to fully explore the 
implications of this figure. 

Nevertheless, we can speculate on a possible interpretation.
Given the location of M54 as the nuclear star cluster of the Sagittarius dwarf galaxy, and given the 
frequently assumed status of $\omega$~Cen as the nuclear remnant of a disrupted dwarf galaxy, it is
tempting to postulate, as did \citet{Sa12}, that the other clusters, including NGC~5824, with significant 
intrinsic [Fe/H] dispersions, i.e., exceeding $\sigma_{int}$([Fe/H]) $\approx$ 0.05 dex, are also the remnant 
central star clusters of tidally disrupted dwarf galaxies.
\citet{BY12} investigated this possibility in the context of NGC~1851
\citep[see also][]{CB10a}, and in the same sense NGC~3201, with its highly retrograde orbit \citep{CD07} 
has long been suggested as having originated in an extra-galactic system accreted by the Milky Way 
\citep[e.g.,][]{RP84}.  As regards $\omega$~Cen, M22 and NGC~5824, the steep slope of the metal-poor 
side of the metallicity distribution is consistent with the rapid enrichment that would be expected to occur 
in the central regions of a dwarf galaxy during its initial formation.  Further evidence to support our 
speculation lies in the discovery of \citet[][see also \citet{CBB12}]{Oz09} that NGC~1851 is surrounded by a
diffuse envelope of stars whose location in the CMD replicate the cluster main sequence.  
The diameter of the NGC~1851 envelope is
$\sim$500 pc, much larger than the nominal tidal radius of the cluster \citep{Oz09, CBB12}.
\citet{Oz09} suggest that the envelope may represent stars lost from the cluster via a variety of dynamical
processes, but it could equally represent the remnants of a disrupted dwarf galaxy in which NGC~1851 was
(or is) embedded \citep{Oz09}.  The situation, however, may be more complex as the spectroscopic 
follow-up of \citet{So12} apparently finds evidence not only for a population of NGC~1851 stars beyond the 
cluster tidal radius, but also for what appears to be an additional stellar stream at a different velocity to that of the 
extended cluster population.

As noted in the Introduction, \citet{HN09} have suggested a connection between NGC~5824 and the Cetus
Polar Stream, which has a similar low metallicity to that of the cluster \citep{HN09}.  Regardless of the 
validity of this possible connection, it is intriguing in this context to note that NGC~5824 is apparently also 
{\it surrounded by a large diffuse stellar envelope}.  The density profile of the cluster as presented 
in \citet{CG95} shows no signs of a tidal cutoff -- rather there is a power-law projected density profile that 
extends well past the nominal tidal radius of $\sim$5.7\arcmin \citep{H96}.  The power-law slope is 
$r^{-2.2\pm0.1}$ and cluster stars are detectable to $log(r\arcmin)$ $\approx$ 1.65 or $r$ $\approx$ 45$
\arcmin$ \citep{CG95}.  \citet{CBB12} find similar results.  At the 32 kpc distance of NGC~5824 \citep{H96}, 
a radius of 45$\arcmin$ corresponds to 420 pc or a diameter approaching one kpc in size.  This is almost 
twice the size of the diffuse envelope surrounding NGC~1851 and approaches the scales of low luminosity
dwarf galaxies.  For example, Bootes~I stars are detected to a radius of $\sim$500 pc \citep{NM08}.  The
\citet{CG95} data also show some indication that the NGC~5824 diffuse envelope is more extended in 
the N-S direction than it is in the E-W direction.  
Given the likely mass of the cluster ($\sim$6$\times10^6 M_{\sun}$ for $M/L_{V}$ = 2)
and the large Galactocentric distance, it is unlikely that tidal effects are responsible for the outer diffuse
envelope of NGC~5824, though an origin with internal dynamical processes remains possible, as does the
interpretation that the envelope represents a remnant population of a disrupted dwarf galaxy.  
A deep large area study of this cluster and its surrounds is clearly called for, and is being undertaken by our 
group.  A spectroscopic survey to determine the mean metallicity and metallicity spread of the stars
in the diffuse envelope would also be worthwhile as it would constrain any differences between the 
abundance properties of the putative nuclear star cluster (NGC~5824), where nucleosynthesis processes 
may have proceeded more quickly, and those of the postulated dwarf galaxy `field star' population.

The results of \citet{LG07} are also relevant to the discussion \citep[see also \S5.3 of][]{Sa12}.  \citet{LG07}
investigated the luminosity distribution and kinematic properties of groups of Galactic globular
clusters selected by HB type.   Based on a total sample of 114 objects, they found 
that the 28 clusters with extended horizontal branches (EHB clusters) are notably more luminous than the
rest of the sample.  The EHB morphology, i.e., a horizontal branch morphology extending 
to very high temperatures, is thought to be associated with the presence of a cluster stellar population that
possesses an enhanced helium abundance \citep[e.g.,][]{DA05}.  Further, \citet{LG07} showed that the EHB 
clusters as a group have kinematics that are dominated by random motions lacking any correlation with 
metallicity.  \citet{LG07} then use these results to argue that
the EHB clusters are a distinct population with an accretion-event origin for most, if not all of the clusters.
As regards our sample of  six clusters with significant intrinsic [Fe/H] dispersions, three 
($\omega$~Cen, M54 and M22) belong to the group with strongly extended horizontal branch 
morphologies, while two (NGC~1851 and NGC~5824) are in the ``moderately extended HB'' group \citep{LG07}.
Only NGC~3201 is not a member of the EHB group.  Consequently, our contention that the clusters with 
significant [Fe/H] variations are the remnant central star clusters of accreted dwarf galaxies is quite
consistent with the \citet{LG07} results.

We end with one further point and that is an estimate of how many former dwarf galaxy nuclear star clusters,
now seen as globular clusters with significant iron abundance intrinsic spreads, might be expected in the 
halo of the Milky Way, given that in our speculation four additional candidates beyond M54 and 
$\omega$~Cen are already identified.  If we assume that the absolute visual magnitude of the Galactic
halo is M$_{V}$ $\approx$ --17 \citep{KCF93} and that of order 50\% of this luminosity is generated from the
disruption of satellite galaxies with the remainder formed in-situ \citep[e.g.,][]{IBJ12}, 
and that the dominant contributors to the
disrupted satellite population have typical luminosities comparable to the Fornax and Sagittarius dwarfs
(i.e., M$_{V}$ $\approx$ --14 \citep{AMC12}), then the disruption of only $\sim$15 or so such systems is implied.
Given that we do not expect all dwarfs of this luminosity to have nuclear star clusters, it is clear that for our 
speculation to be valid 
there should not be more than a few additional Galactic globular clusters with significant intrinsic iron
abundance spreads awaiting discovery.   \citet{H96} lists 36 clusters with M$_{V}$ $\leq$ --8; of these 12
are included in Fig.\ \ref{sprd_vs_Mv_fig}.  Of the remainder, some are ``standard clusters'' such as 
M92 and M3 for which there is no evidence for the presence of significant intrinsic iron abundance spreads.  
Nevertheless, there does remain a number of relatively unstudied clusters in this group.  We 
predict future work will reveal the presence of intrinsic [Fe/H] spreads in a handful of these systems at most, 
else our speculation will have to be revisited.

In summary, we have used Ca II triplet spectroscopy of a large and well-defined sample of RGB members
to establish that the luminous outer-halo globular cluster NGC~5824 possesses an intrinsic [Fe/H] dispersion. 
The dispersion is
characterised by an inter-quartile range of 0.10 dex and a total abundance range of $\sim$0.3 dex.  NGC~5824
thus joins a small number of other Galactic globular clusters with intrinsic [Fe/H] abundance dispersions.  As for
$\omega$~Cen and M22, the abundance distribution rises steeply on the metal-poor side.  The cluster is also
apparently contained within an extended stellar envelope of $\la$1 kpc in size.  Together these properties
suggest that NGC~5824 may be the remnant nuclear star cluster of a former dwarf galaxy accreted and 
disrupted by the Milky Way. 

\section*{Acknowledgements}

G.\ Da C.\ would like to acknowledge research support from the Australian Research Council through
Discovery Grant programs DP120101237 and DP120100475.  He is also grateful for the support 
received during an extended visit to the Institute of Astronomy, University of Cambridge, and for shorter
visits to Osservatorio Astronomico di Padova, INAF and ESO-Santiago. 

Based in part on observations 
for program GS-2011A-Q-47 obtained at the Gemini Observatory, which is operated by the 
Association of Universities for Research in Astronomy, Inc., under a cooperative agreement 
with the NSF on behalf of the Gemini partnership: the National Science Foundation (United 
States), the Science and Technology Facilities Council (United Kingdom), the 
National Research Council (Canada), CONICYT (Chile), the Australian Research Council (Australia), 
Minist\'{e}rio da Ci\^{e}ncia, Tecnologia e Inova\c{c}\~{a}o (Brazil) 
and Ministerio de Ciencia, Tecnolog\'{\i}a e Innovaci\'{o}n Productiva (Argentina).

Based also in part on observations obtained under ESO program 087.D-0465.


\end{document}